\documentclass[sigconf]{acmart}

\AtBeginDocument{%
  }

\copyrightyear{2025}
\acmYear{2025}
\setcopyright{acmlicensed}
\acmConference[MM '25] {Proceedings of the 33rd ACM International Conference on Multimedia}{October 27--31, 2025}{Dublin, Ireland.}
\acmBooktitle{Proceedings of the 33rd ACM International Conference on Multimedia (MM '25), October 27--31, 2025, Dublin, Ireland}
\acmISBN{979-8-4007-2035-2/2025/10}
\acmDOI{10.1145/3746027.3758186}

\usepackage{subcaption}
\usepackage{float}
\usepackage{balance}

\begin{document}

\title{Teaching AI to Feel: A Collaborative, Full-Body Exploration of Emotive Communication}

\author{Esen K. Tütüncü}
\orcid{0000-0002-0050-0908}
\affiliation{%
  \institution{Institute of Neurosciences of the University of}
  \city{Barcelona}
  \country{Spain}
}
\email{esenkucuktutuncu@ub.edu}

\author{Lissette Lemus}
\affiliation{%
  \institution{Artificial Intelligence Research Institute (IIIA-CSIC)}
   \city{Barcelona}
  \country{Spain}}

\author{Kris Pilcher}
\affiliation{%
  \institution{Massachusetts Institute of Technology}
  \city{Cambridge}
  \country{USA}
}

\author{Holger Sprengel}
\affiliation{%
 \institution{ESPRONCEDA Institute of 
\newline Art \& Culture}
 \city{Barcelona}
 \country{Spain}}

\author{Jordi Sabater-Mir}
\affiliation{%
  \institution{Artificial Intelligence Research Institute (IIIA-CSIC)}
   \city{Barcelona}
  \country{Spain}}

\renewcommand{\shortauthors}{K. Tütüncü et al.}


\begin{abstract}

Commonaiverse is an interactive installation exploring human emotions through full-body motion tracking and real-time AI feedback. Participants engage in three phases: Teaching, Exploration and the Cosmos Phase, collaboratively expressing and interpreting emotions with the system. The installation integrates MoveNet for precise motion tracking and a multi-recommender AI system to analyze emotional states dynamically, responding with adaptive audiovisual outputs. By shifting from top-down emotion classification to participant-driven, culturally diverse definitions, we highlight new pathways for inclusive, ethical affective computing. We discuss how this collaborative, out-of-the-box approach pushes multimedia research beyond single-user facial analysis toward a more embodied, co-created paradigm of emotional AI. Furthermore, we reflect on how this reimagined framework fosters user agency, reduces bias, and opens avenues for advanced interactive applications. 
\end{abstract}

\ccsdesc[500]{Human-centered computing~Collaborative interaction}
\ccsdesc[300]{Human-centered computing~HCI theory, concepts and models}
\ccsdesc[300]{Applied computing~Media arts}
\ccsdesc[300]{Computing methodologies~Artificial intelligence}

\ccsdesc[500]{Human-centered computing~HCI theory, concepts and models}

\keywords{Affective Computing, Full-Body Emotion Recognition, Participatory AI, Real-Time Feedback Systems, Embodied Interaction}

\begin{teaserfigure}
\centering
  \includegraphics[width=1\textwidth]{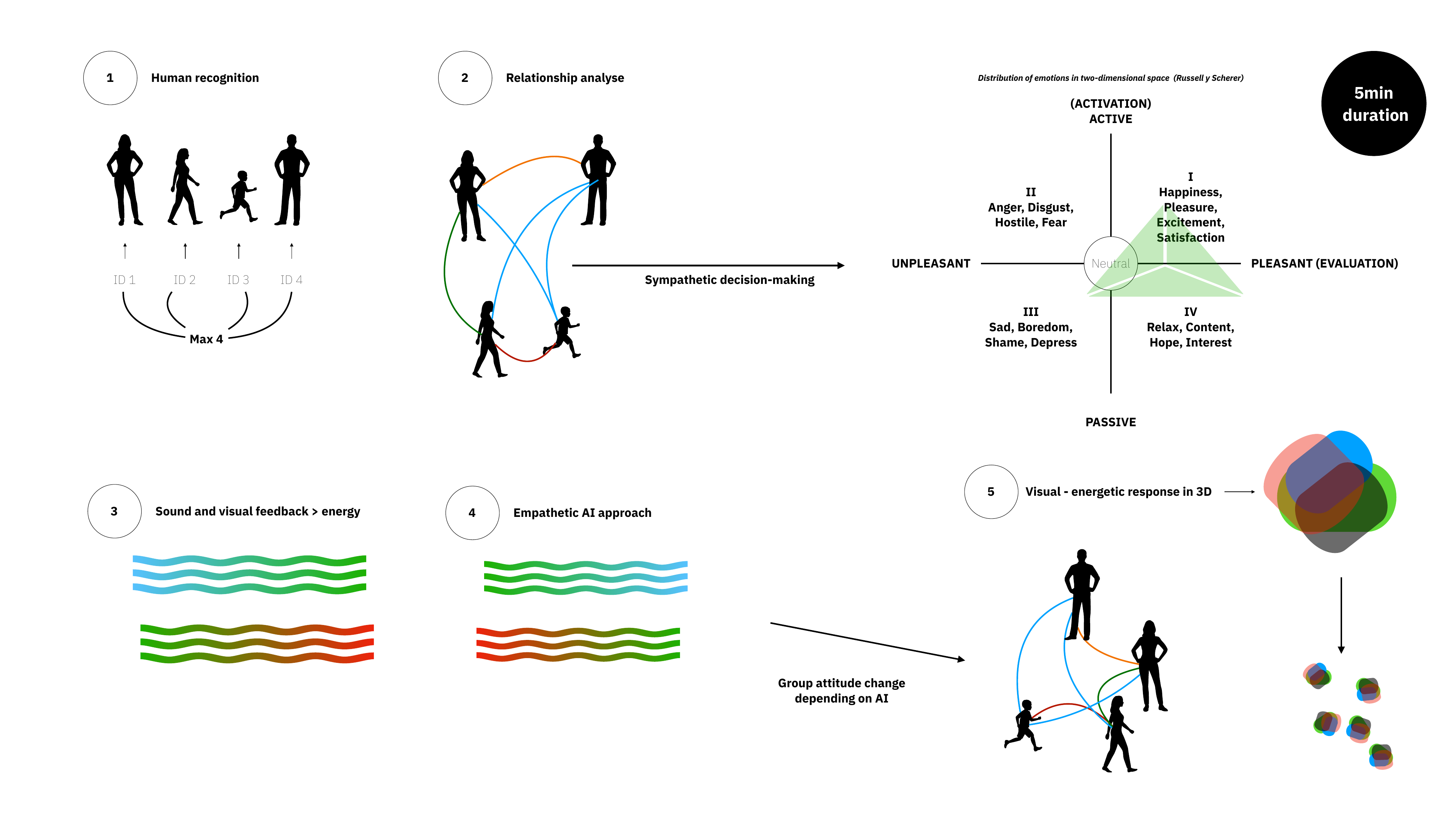}
    \caption{The interaction flow: 1) The AI agent detects the humans inside the room. 2) Assessment of both individual and collective motions and group dynamics 3) The level of energy and movement serving as a metric for the amplitude of the measured emotion 4) Prediction of emotions based on user data 5) Real-time visualization of the perceived emotions.}
  \Description{An interaction flow visualization of the installation. 1) Group of people with various age and gender. 2) The dynamics of the relationships shown with connecting lines. 3) The emotional distribution in 2 axis visualization based on Russel\&Scherer followed by the emotion visualization through the group interactions. 4) Abstraction of emotions proposed by AI 5) Visualization of emotion through various morphing forms.}
  \label{fig:interactionflow}
\end{teaserfigure}


\maketitle

\section{Introduction: A Collective Embodiment of Emotion}
Artificial intelligence (AI) systems are increasingly tasked with interpreting human emotions, with much of the field historically rooted in exploratory but reductionist frameworks such as Ekman’s Facial Action Coding System \cite{ekman1978facial}. This influential work proposed that universal categories of emotions could be tied to specific facial expressions, offering a seemingly intuitive pathway for decoding human affect. The appeal of facial expression analysis lies in its simplicity: the face, as a visible and universally present interface of emotion, provides a natural focal point for researchers aiming to systematize emotional understanding. Studies built upon Ekman’s framework have driven significant advancements in emotion recognition technologies, enabling applications in diverse domains such as advertising, health monitoring, and workplace productivity \cite{mcstay2018emotional}.
Despite its influence, this paradigm has faced growing criticism for oversimplifying the multifaceted nature of emotions. The assumption that emotions are biologically hardwired and universally expressed has been challenged by scholars such as Barrett and Russell, who argue that emotions are constructed phenomena shaped by context, personal experience, and cultural norms\cite{russell1994there,barrett2014psychological}. These critiques expose the limitations of face-centric models, which often fail to capture the complexity, fluidity, and socially situated nature of emotional expression.	

Affective computing, since its inception, has largely been dominated by facial and vocal analysis, employing methods such as Support Vector Machines (SVMs), Hidden Markov Models (HMMs), and, more recently, Convolutional Neural Networks (CNNs) \cite{tzirakis2017end}. While these techniques have demonstrated success in analyzing facial images, vocal tones, and textual sentiment, they often rely on static datasets that fail to account for the dynamic, context-dependent nature of emotional communication. Moreover, the emphasis on facial expressions risks perpetuating a narrow and fragmented understanding of emotions, particularly in social and embodied contexts \cite{picard2000affective}.


Another critical limitation of existing systems—particularly in multimedia research contexts—lies in their neglect of full-body movements and gestures. While facial expressions and vocal tones play a significant role in emotional communication, research has shown that body language often conveys affective states more effectively than the face alone \cite{schmidt2001human,atkinson2007evidence}. Gestures, postures, and movement patterns provide rich, multimodal cues that can enhance emotional interpretation, yet these remain underrepresented in current AI frameworks \cite{kleinsmith2012affective}. Although multimodal approaches integrating facial, vocal, and textual data have shown promise \cite{poria2018multimodal}, the gap in incorporating dynamic, whole-body movements remains a significant barrier to achieving a holistic understanding of emotion in interactive multimedia environments.

Beyond the limitations of modality, a critical shortcoming of current emotion recognition systems is their inability to account for the nuanced cultural and social frameworks that shape emotional expression. Emotions are deeply influenced by interpersonal interactions and societal norms, which vary widely across different communities and contexts. For instance, the way joy or sadness is expressed can diverge significantly across cultures, reflecting variations in norms, traditions, and values \cite{elfenbein2002universality}. Studies such as those by Gendron et al. have shown that interpretations of facial expressions differ depending on cultural perspectives, emphasizing the need for AI systems to move beyond assumptions of universal emotional standards \cite{gendron2014perceptions}. Addressing these gaps requires AI models that can adapt dynamically to diverse emotional landscapes, integrating localized and context-sensitive insights to ensure inclusivity and relevance.

Compounding these issues is the reliance on predefined emotion categories, which oversimplifies the fluid and often overlapping nature of human affect. Emotions exist on a spectrum and can be experienced simultaneously, making their classification into discrete categories inherently reductive. As mentioned before, Barrett and Russell advocate for a constructionist view of emotions, emphasizing their emergent and context-dependent nature \cite{barrett2014psychological}. Similarly, Sherry Turkle critiques the growing trend of quantifying emotions into binary or discrete values, warning that such approaches risk prioritizing convenience over depth and fostering superficial connections \cite{turkle2017alone}.	

Ethical considerations further complicate the deployment of AI emotion recognition systems. Crawford and Calo caution that reductionist models often fail to address biases inherent in their datasets, reinforcing systemic inequalities and overlooking privacy concerns \cite{crawford2016there}. The emphasis on facial expressions, for example, has been shown to carry racial biases, further marginalizing underrepresented groups. These limitations underscore the need for AI systems that move beyond static datasets, predefined labels, and narrow modalities to engage with emotions as embodied, dynamic, and culturally situated phenomena.

In light of these challenges, the question arises: How can AI move beyond its current limitations to engage with emotions in ways that are more inclusive, adaptive, and reflective of their embodied nature? To explore this question, we present Commonaiverse, an interactive multimedia installation that enables individuals to express and interpret human emotions through full-body movements, exploring the dynamic relationship between humans and AI in real-time. By emphasizing participatory and multisensory engagement, Commonaiverse seeks to reframe how emotions are understood, not as fixed categories but as living, evolving expressions grounded in social and cultural contexts.
\section{Designing the Commonaiverse}
The Commonaiverse installation was conceived as a space to explore and express emotions through full-body movements, encouraging participants to reconnect with the physicality of their emotions while acknowledging the interconnectedness of human experience. In designing this system, we sought to challenge the increasingly isolated and digitized modes of expression in modern life by fostering collaboration and mutual engagement. The installation deliberately requires at least two participants, emphasizing the inherently social and interdependent nature of emotional communication.
\subsection{Conceptual Framework}
The design of Commonaiverse is based in the understanding that emotions emerge through interactions,both with others and the environment, and are deeply tied to physical movement. By emphasizing full-body gestures, postures, and dynamic movements, the installation fosters a multisensory exploration of emotions, bridging the gap between abstract data representations and lived, embodied experiences. This contrasts sharply with the disembodied focus of traditional AI emotion recognition systems, which often reduce affective states to facial or vocal features alone.
\begin{figure}[ht]
  \centering
  \includegraphics[width=1\linewidth]{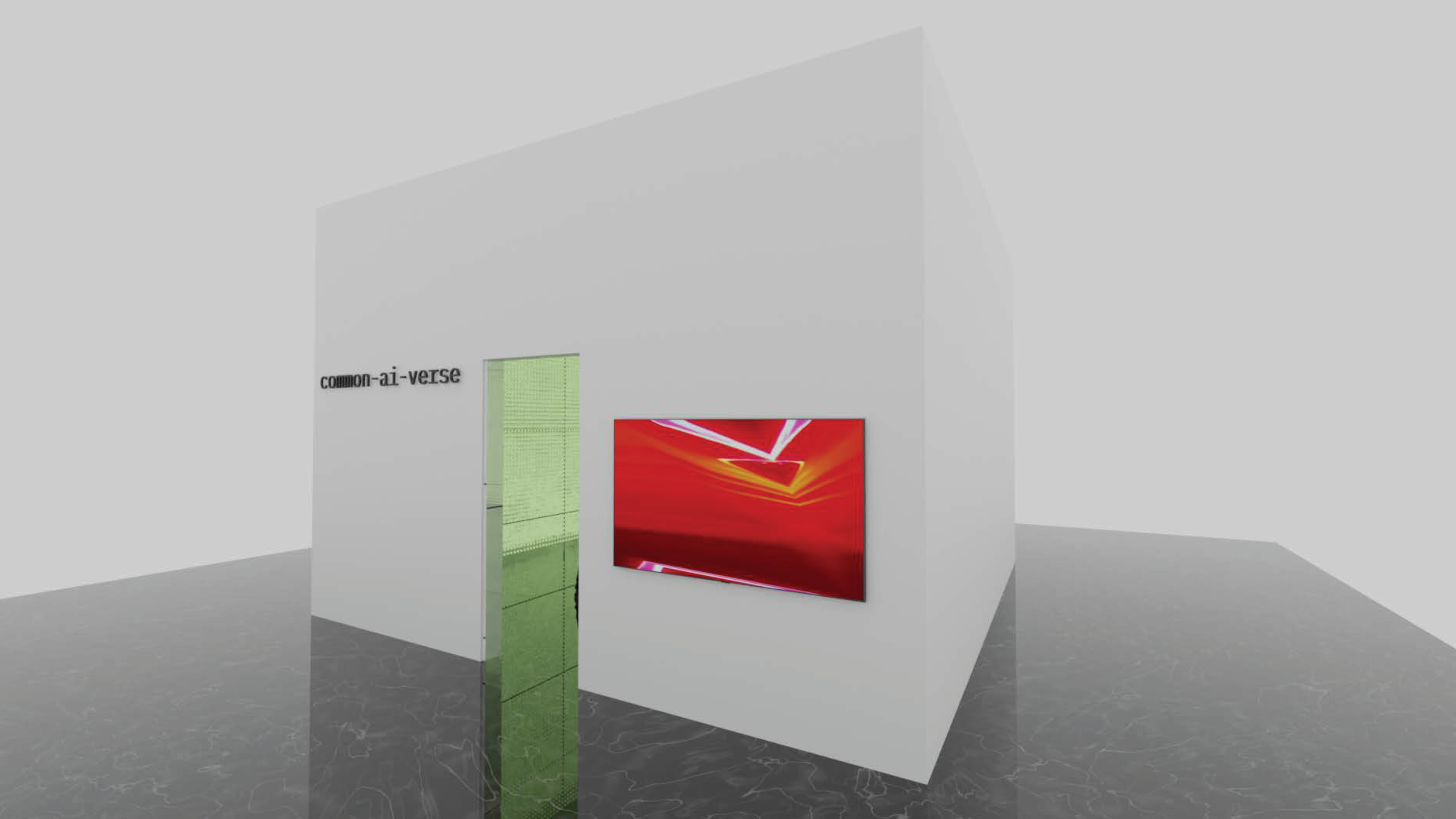}
  \caption{Early model for the space, where the walls serve as a container for the interaction.}
  \Description{A white cube-like room where the interaction happens, and a screen on the outside.}
\label{fig:fig2}
\end{figure}

\begin{figure}[h]
  \centering
  \includegraphics[width=1\linewidth]{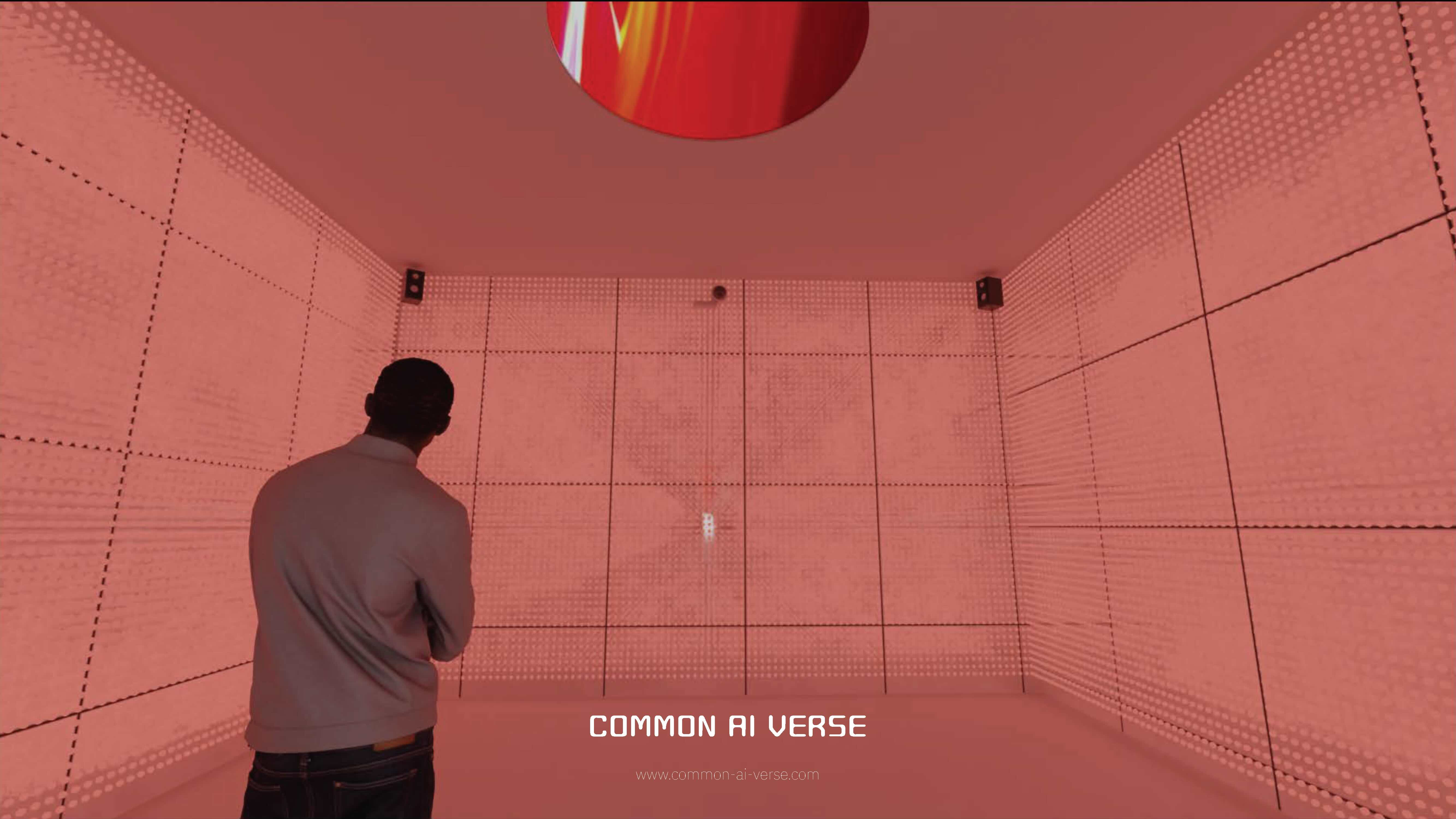}
  \caption{3D Model where the ceiling was also thought to be a part of the visual narrative.}
  \Description{Inside of the room where the users interact, surrounded by led walls.}
  \label{fig:fig2-1}
\end{figure}

To ensure participants feel comfortable expressing themselves, the space was designed as an enclosed, private environment, shielding individuals from external judgment or distractions. This enclosed design, illustrated in Figures \ref{fig:fig2} and \ref{fig:fig2-1}, encourages authentic, uninhibited participation by providing a safe setting for emotional exploration. The room itself, with its integrated visual and spatial elements, supports the fluid exchange of energy and interaction between participants, reinforcing the relational nature of emotions while creating a contained yet immersive narrative environment.
\subsection{Early Iterations}
\begin{figure}[ht]
  \centering
  \includegraphics[width=0.7\linewidth]{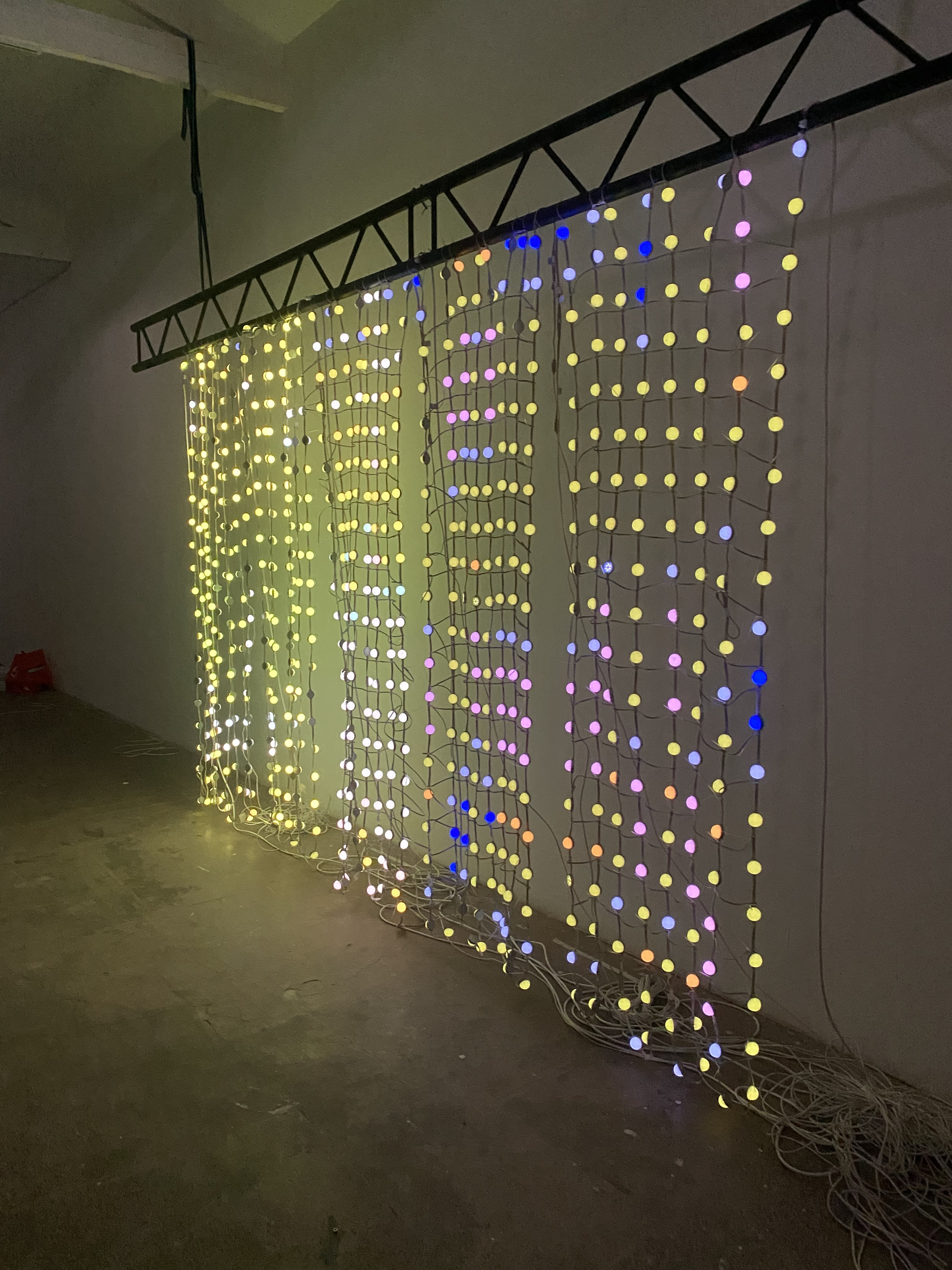}
  \caption{Although entertaining, having the ledwalls meant the users would focus more on the visuals than the interaction that is taking place}
  \Description{Led strips formed into a wall where the visuals take place}
\label{fig:ledwall}
\end{figure}

Initial sketches and prototypes focused on creating a reactive environment using LED light strips with low-resolution grids placed along the walls (see Figure \ref{fig:ledwall}). These early designs mapped participants' movements onto the grid, using shifting colors and patterns to represent their emotional states in real time. However, observations during testing revealed that these visualizations often distracted participants, drawing their attention to the walls rather than fostering connection with their co-participants or encouraging self-expression.
\subsection{Refinement of the Interactive Space}
\begin{figure}[h]
  \centering
  \includegraphics[width=1\linewidth]{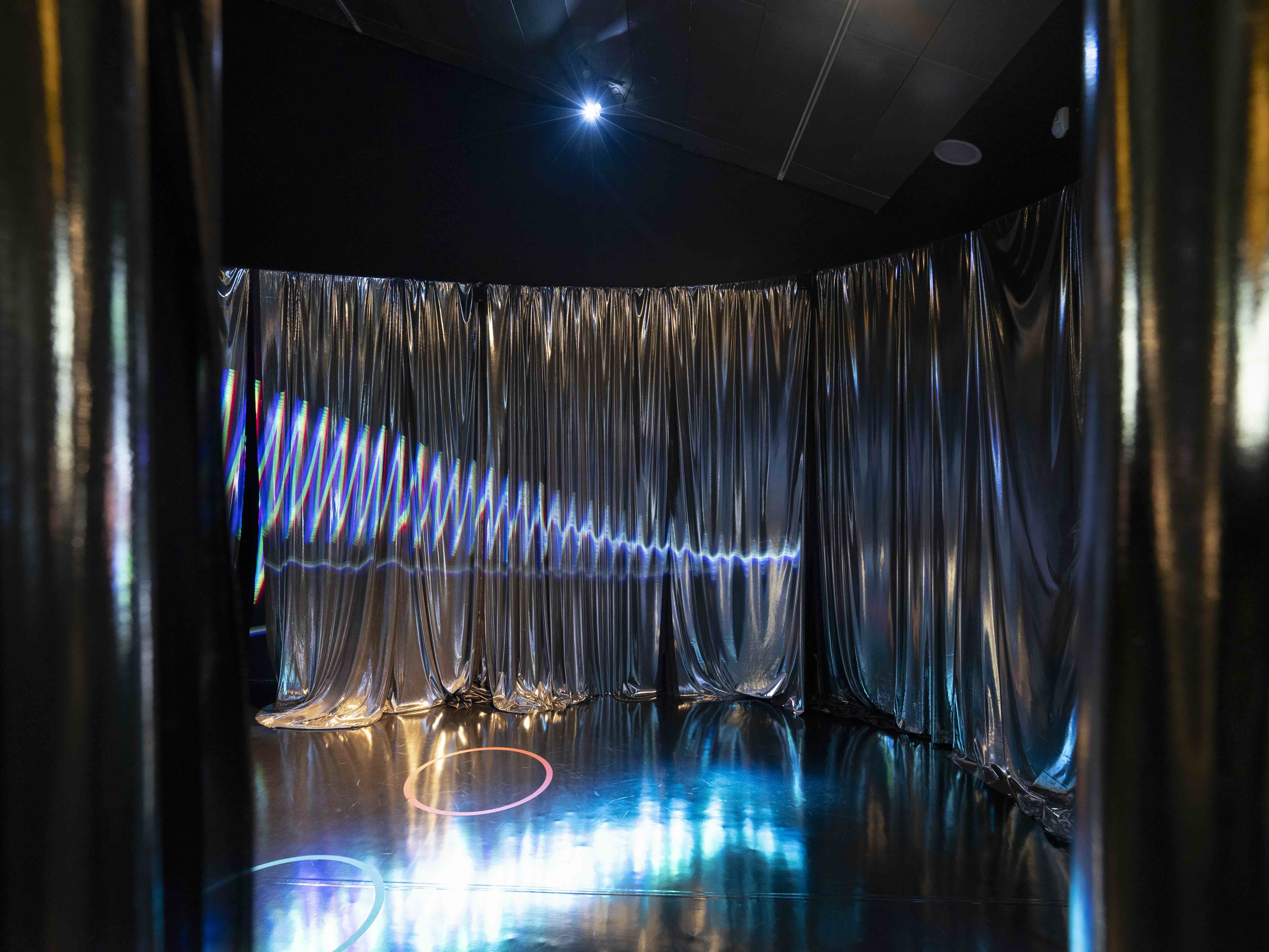}
  \caption{Final configuration of the room, with lasers projected on the reflective fabric, creating ripple effects.}
  \Description{Room surrounded by shiny fabric with lasers projected on them}
\label{fig:laserinside}
\end{figure}

Recognizing this limitation, we shifted the focus of the design to prioritize the participants’ bodily interactions and the AI’s responses rather than emphasizing external visual feedback. The final design minimized reliance on wall-based visuals and instead introduced dynamic, abstract audiovisual outputs projected into the space and the sheer fabric walls via three lasers mounted on the ceiling, which can be seen in Figure \ref{fig:laserinside}. This decision allowed the participants to remain present in the shared experience rather than disengaging to observe the system’s reactions.

\subsection{Interaction Flow}
The interaction within Commonaiverse unfolds in three sequential phases, highlighted in Figure \ref{fig:interactionflow}, with the entire experience lasting between 15–20 minutes. Each phase varies in duration, typically lasting 5–7 minutes, and is designed to deepen the participants’ engagement with the system and one another:
\begin{itemize}
\item \textbf{Teaching Phase:}
Participants begin by demonstrating specific emotional states, such as sadness, joy, or anger, using full-body movements. For instance, a participant might slowly crouch and hunch their shoulders to express sadness or leap with outstretched arms to convey joy. These movements are captured and analyzed in real-time.
The AI system learns from these movements, creating a data map that correlates body gestures with emotional labels. This phase emphasizes active teaching, where participants co-create the emotional lexicon with the AI rather than working from a pre-existing dataset.

\item \textbf{Exploration Phase:}
In this phase, participants are invited to perform unscripted, free-form movements. The agent attempts to interpret their emotions based on the patterns it has learned during the teaching phase. This introduces a dynamic feedback loop, where participants can evaluate and respond to the AI’s interpretations.
The system’s responses are displayed as abstract audiovisual outputs, evolving in real time to reflect the emotional interplay between participants and the AI.
\item \textbf{Emotional Cosmos Phase:}
The culmination of the interaction is represented as a digital “cosmos” that visualizes the emotional exchanges that occurred during the session. This visualization is displayed on an external panel, also made accessible via a unique QR code generated for each session to retrieve afterwards. This phase provides participants with a lasting representation of their engagement, highlighting the relational and evolving nature of emotions.
\end{itemize}

\section{Implementation}
\subsection{Body Tracking}
For the body tracking we used MoveNet, a state-of-the-art human pose estimation model, for real-time tracking of participants’ full-body movements. MoveNet was selected due to its demonstrated robustness and high accuracy compared to other models in both controlled and real-world environments, as highlighted in recent evaluations \cite{jo2022comparative, goyal2023moveenet}. Its lightweight architecture allows for efficient operation on edge devices, ensuring smooth interaction without latency, which is critical for maintaining participant immersion.

MoveNet operates by estimating 17 key body points, including major joints and body landmarks, with high temporal precision. The system is optimized to handle variations in body orientation, lighting conditions, and occlusions, which are common in interactive installations. Data streams from MoveNet were processed through custom Python scripts, using the TensorFlow Lite framework for compatibility. These scripts were responsible for configuring the model, interpreting keypoint data, and transmitting pose estimations to the multimedia components of the installation via Open Sound Control (OSC).

To enhance the temporal resolution and ensure robustness against erratic movements, post-processing techniques were applied. These included smoothing algorithms to interpolate noisy frames and dynamic calibration to adapt to variations in participant height and posture. Additionally, a confidence threshold was established for each keypoint to filter out inaccuracies, enabling the system to focus on meaningful gestures and movements.

\begin{figure}[h]
  \centering
  \includegraphics[width=1\linewidth]{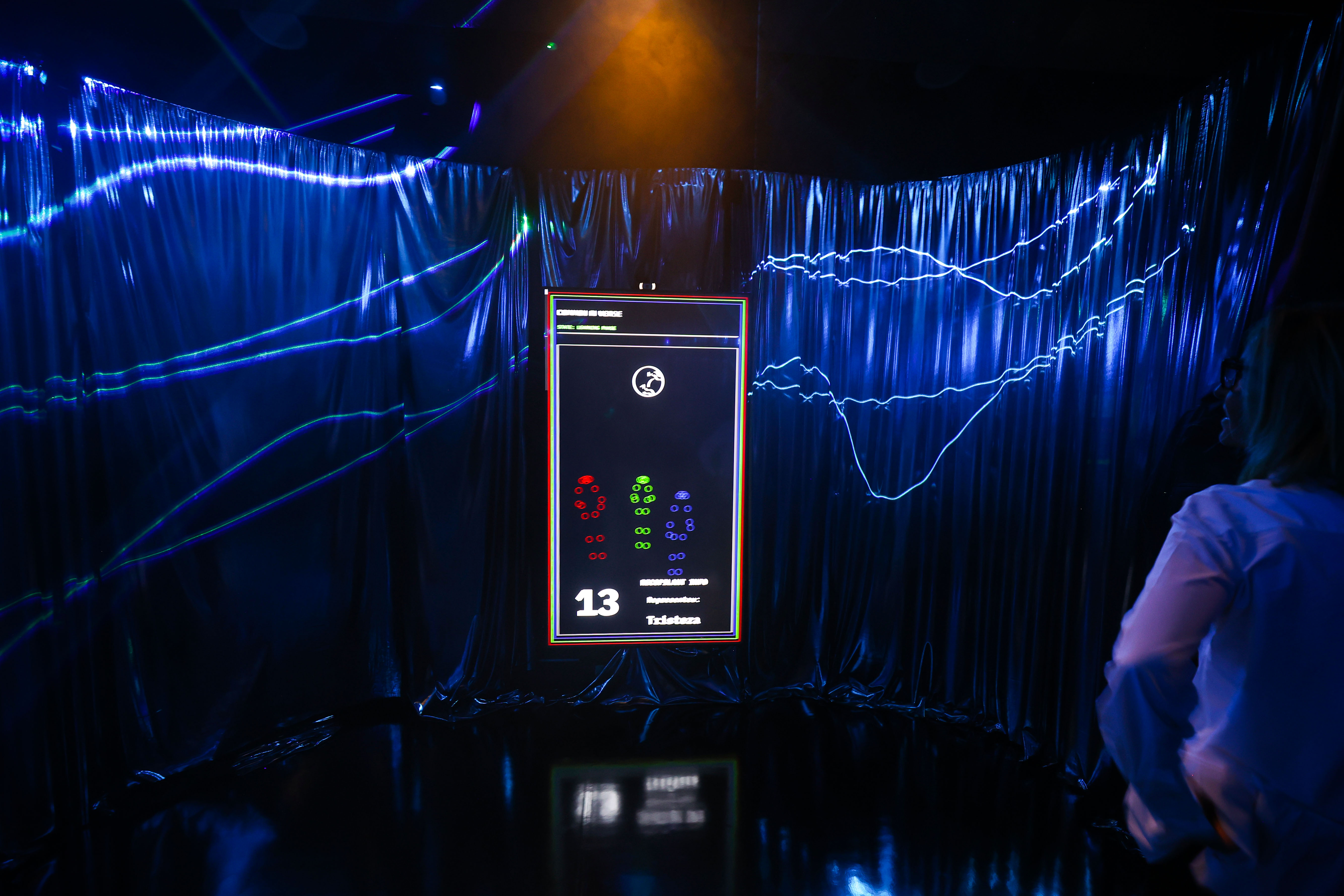}
  \caption{Body Tracking visualization during the session, the users were identified with 3 different (RGB) colors.}
  \Description{The screen inside the room showing the body tracking visualization of the users.}
\end{figure}
\subsection{AI Pipeline}
The AI architecture operates through a structured multi-recommender system \cite{inbook}, analyzing emotional expression in real-time across distinct phases of interaction. The multi-recommender approach relies on specialized entities (recommenders) that analyze data from different perspectives, such as movement amplitude, speed, frequency, and interpersonal proximity, to interpret emotions. Each recommender focuses on specific interaction aspects, providing its own assessment. These inputs from the recommenders are then aggregated to establish a final consensus.This framework ensures that participants’ movements and emotional states are accurately captured, contextualized, and adapted over the course of the installation.

The system is structured into three primary phases, each building upon the preceding one to deepen engagement and refine emotional interpretation:
\begin{itemize}
    \item \textbf{Preparation Phase:} The system initializes by detecting predefined emotions—happiness, relaxation, anger, and sadness—using behavioral metrics such as movement amplitude, speed, and proximity of participants. MoveNet provides the foundational motion data, which is analyzed to establish a baseline emotional state. This phase ensures the system is primed to adapt to dynamic interactions.
    \item \textbf{Detection Phase:} During live interactions, the system iteratively detects and refines emotional states using continuous data streams from MoveNet alongside contextual inputs. The AI dynamically adjusts its predictions to reflect emerging movement patterns and behaviors. By integrating these insights, the system generates audiovisual outputs that align with participants' evolving emotional expressions, ensuring the experience remains responsive and immersive.
    \item \textbf{Evaluation and Adaptation Phase:} Detected emotional states are monitored over time to assess trends or inconsistencies. If shifts in behavior occur, the system reevaluates its predictions and updates its emotional mapping. This phase not only ensures alignment with contextual factors but also provides participants with feedback reflective of their sustained interactions.
\end{itemize}
The multi-recommender system enables granular analysis across these phases, with specific modules tailored to the installation:
\begin{enumerate}
    \item \textbf{Behavioral Data Recommender (REC1):} REC1 extracts movement features such as speed, amplitude, and participant proximity. These metrics form the foundation for identifying emotional states based on physical patterns, providing a direct link between body language and affective states.
    \item \textbf{Contextual Recommender (REC2):} REC2 enhances emotional interpretations by integrating situational variables, including the number of participants and their spatial relationships. By contextualizing movement data, this recommender captures group dynamics and environmental nuances, offering a more comprehensive understanding of collective emotions.
    \item \textbf{Longitudinal Emotion Recommender (REC3):} REC3 tracks emotional trends across time, analyzing how states evolve during a session. This module adds a temporal layer to the AI’s insights, allowing the system to adapt to sustained or shifting emotional expressions dynamically.

\end{enumerate}
While the Commonaiverse installation emphasizes REC1, REC2, and REC3, the architecture also includes provisions for future expansions with additional modules. These modules address challenges associated with traditional systems that rely on pre-made datasets, instead utilizing participant-driven data generated during the experience to ensure adaptability and context-aware emotional interpretations. Specifically:
\begin{enumerate}
    \item \textbf{Facial Expression Recommender (REC4):}  Processes facial cues dynamically, avoiding reliance on static datasets by integrating real-time participant input. This approach prioritizes context-sensitive interpretations that better reflect the nuances of live interactions.
    \item \textbf{Voice Tone Recommender (REC5):} Captures vocal characteristics and maps them to emotional states in conjunction with user-specific inputs. This ensures flexibility and relevance across diverse participants and cultural contexts.
    \item \textbf{Physiological Signal Recommender (REC6):}  Envisions using biometric data such as heart rate or skin conductance to refine emotional arousal detection. Although not currently implemented, this module represents a potential direction for more comprehensive analysis.
\end{enumerate}
The modularity of the multi-recommender system allows for scalability, ensuring that future expansions can integrate diverse data types for richer emotional analysis.
\subsection{Interactive Real-Time Feedback}
The interactive feedback system in Commonaiverse was developed to integrate participants' body movements and emotional states into the installation's audiovisual outputs in real time. Using TouchDesigner \cite{touchdesigner}, the system processed continuous data streams received via OSC from the motion tracking and AI emotion recognition components. These data streams included body position, movement velocity, and inferred emotional states, which were used to manipulate laser visuals projected on reflective fabric walls and update the main display interface visible to participants.

Within TouchDesigner, incoming data was mapped to control parameters of the laser projections, which consisted of geometric patterns and color schemes that dynamically adjusted based on the participants’ interactions. For instance, increased motion intensity or speed influenced the complexity and fluidity of the laser visuals, while shifts in emotional states altered the colors and overall aesthetic tone of the projections. In the meanwhile, the main display provided a real-time visualization of participants' session data, including movement metrics such as quantity, speed, and range of motion, as well as emotion-specific values such as detected levels of "happiness," "relaxation," or "sadness."
\begin{figure}[ht]
  \centering
    \centering
    \includegraphics[width=1\linewidth]{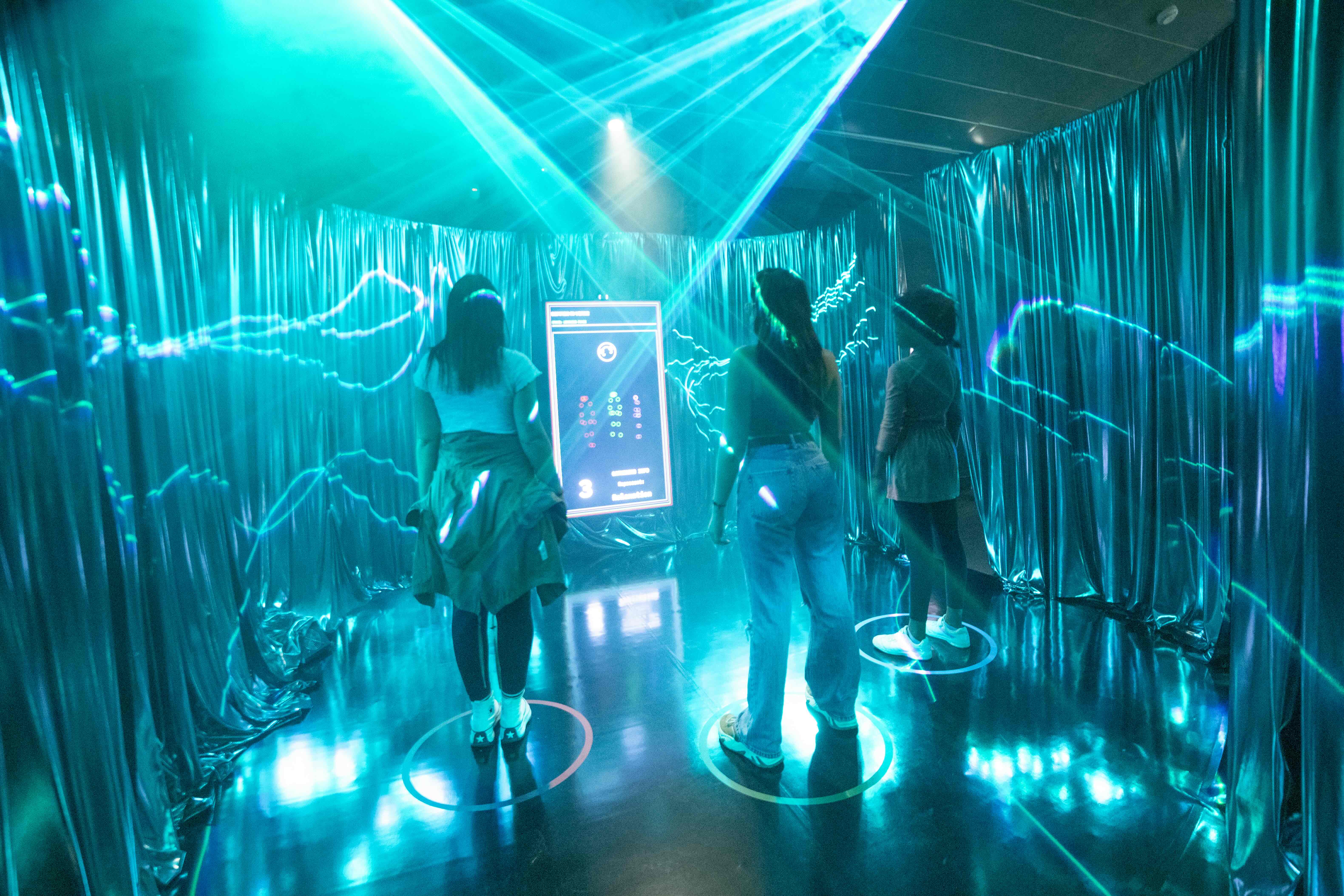}
    \caption{Participants starting the training phase of the Commonaiverse}
  \Description{Three people inside the installation where the blue light shines on them and we can see the body-tracking visualization on the screen}
\end{figure}

This real-time system was designed to provide participants with immediate feedback on their interactions while maintaining consistency between physical movements, interpretation of emotions, and projected outputs. By linking motion data and emotional states directly to the visual and data interfaces, the system created an interaction loop that captured and reflected participants’ contributions in both artistic and measurable ways. The use of TouchDesigner allowed for flexible, real-time adjustments, ensuring the feedback stayed responsive to the evolving dynamics of the interaction.
\begin{figure*}[ht]
  \centering
  \begin{subfigure}[b]{0.22\textwidth}
    \centering
    \includegraphics[width=\linewidth]{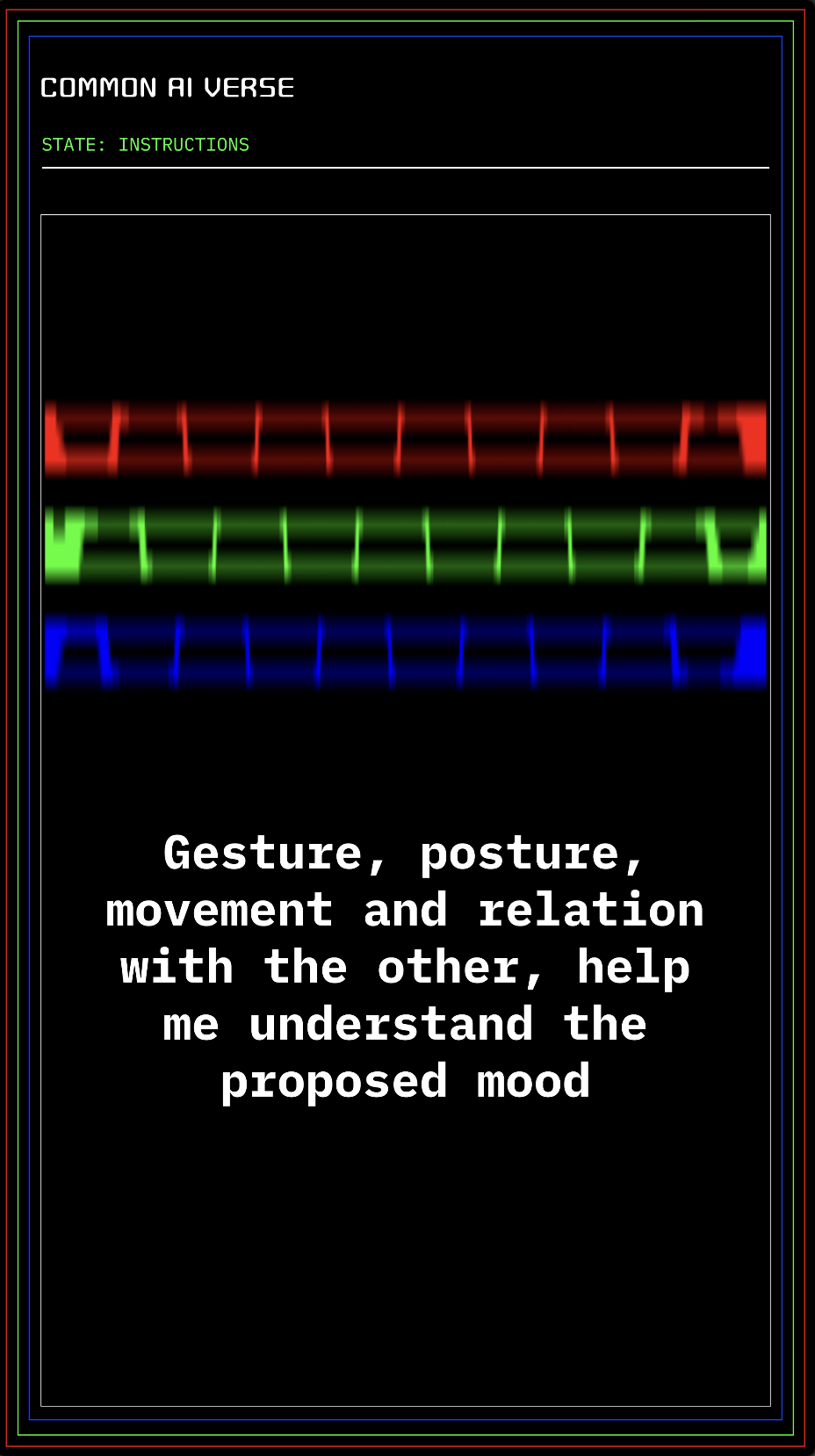}
    \caption{Idle state}
  \end{subfigure}  \hfill
  \begin{subfigure}[b]{0.22\textwidth}
    \centering
    \includegraphics[width=\linewidth]{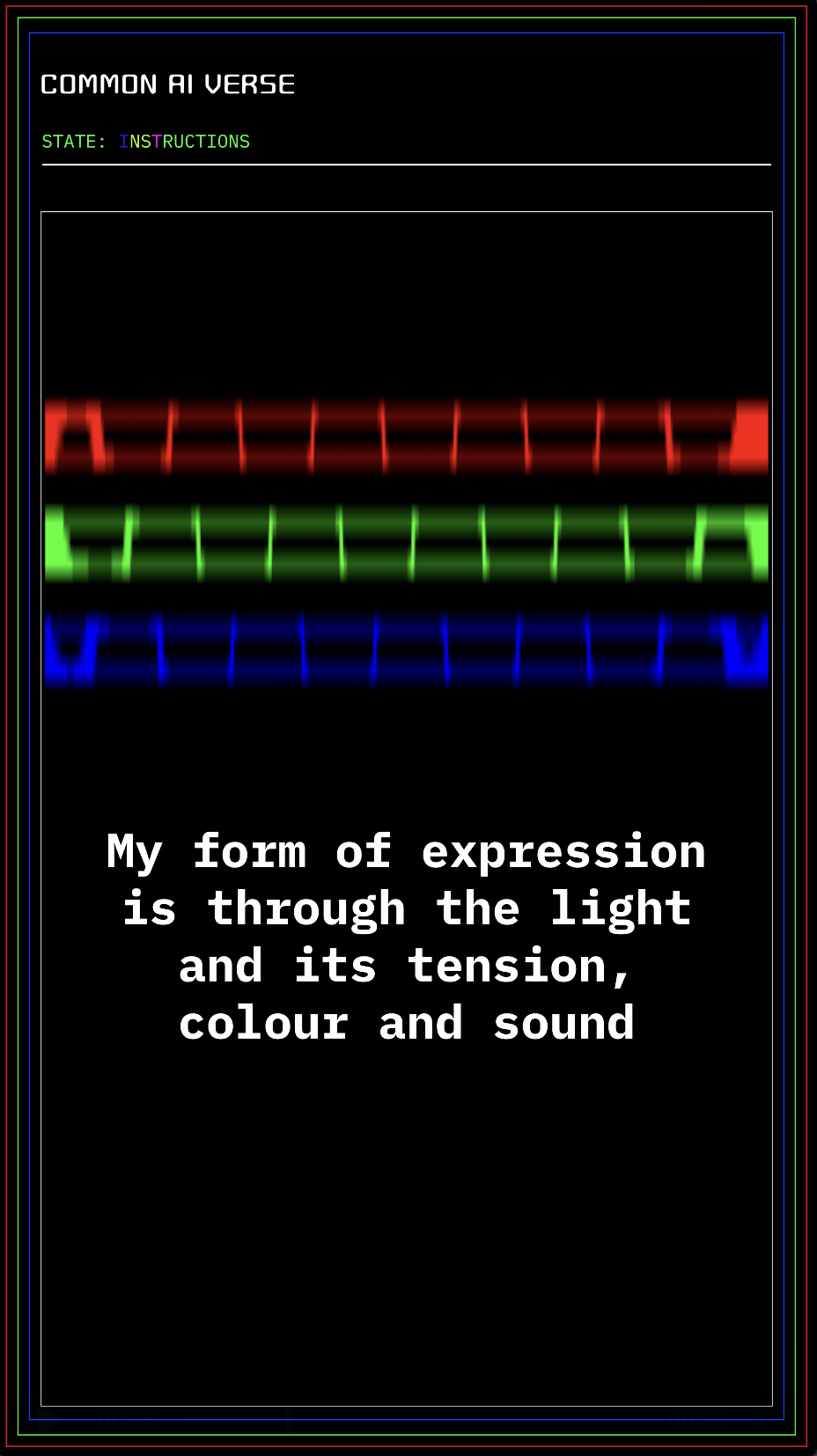}
    \caption{Initializing instructions}
  \end{subfigure}
  \hfill
  \begin{subfigure}[b]{0.22\textwidth}
    \centering
    \includegraphics[width=\linewidth]{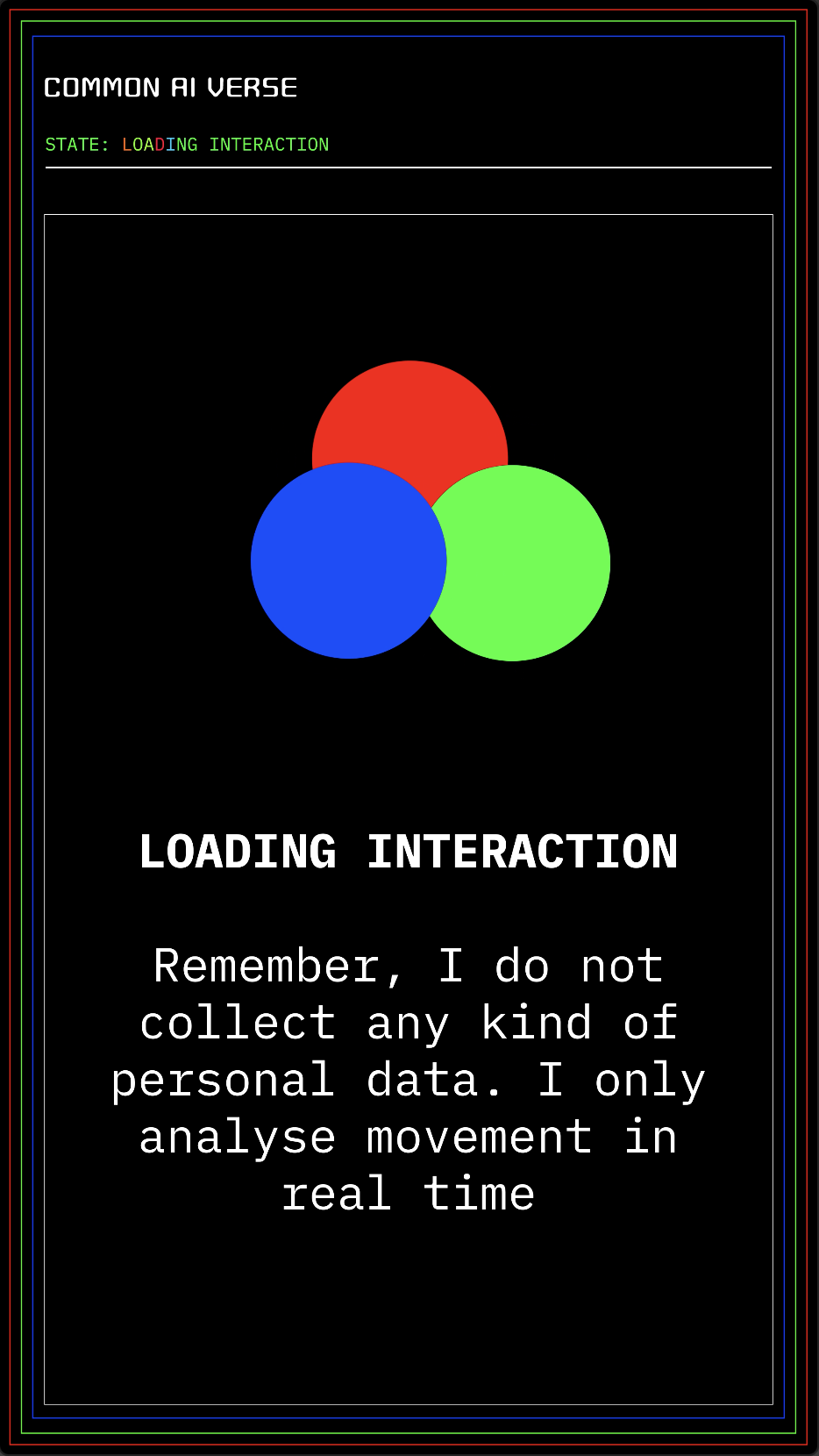}
    \caption{Loading the interaction}
  \end{subfigure}
  \hfill
  \begin{subfigure}[b]{0.22\textwidth}
    \centering
    \includegraphics[width=\linewidth]{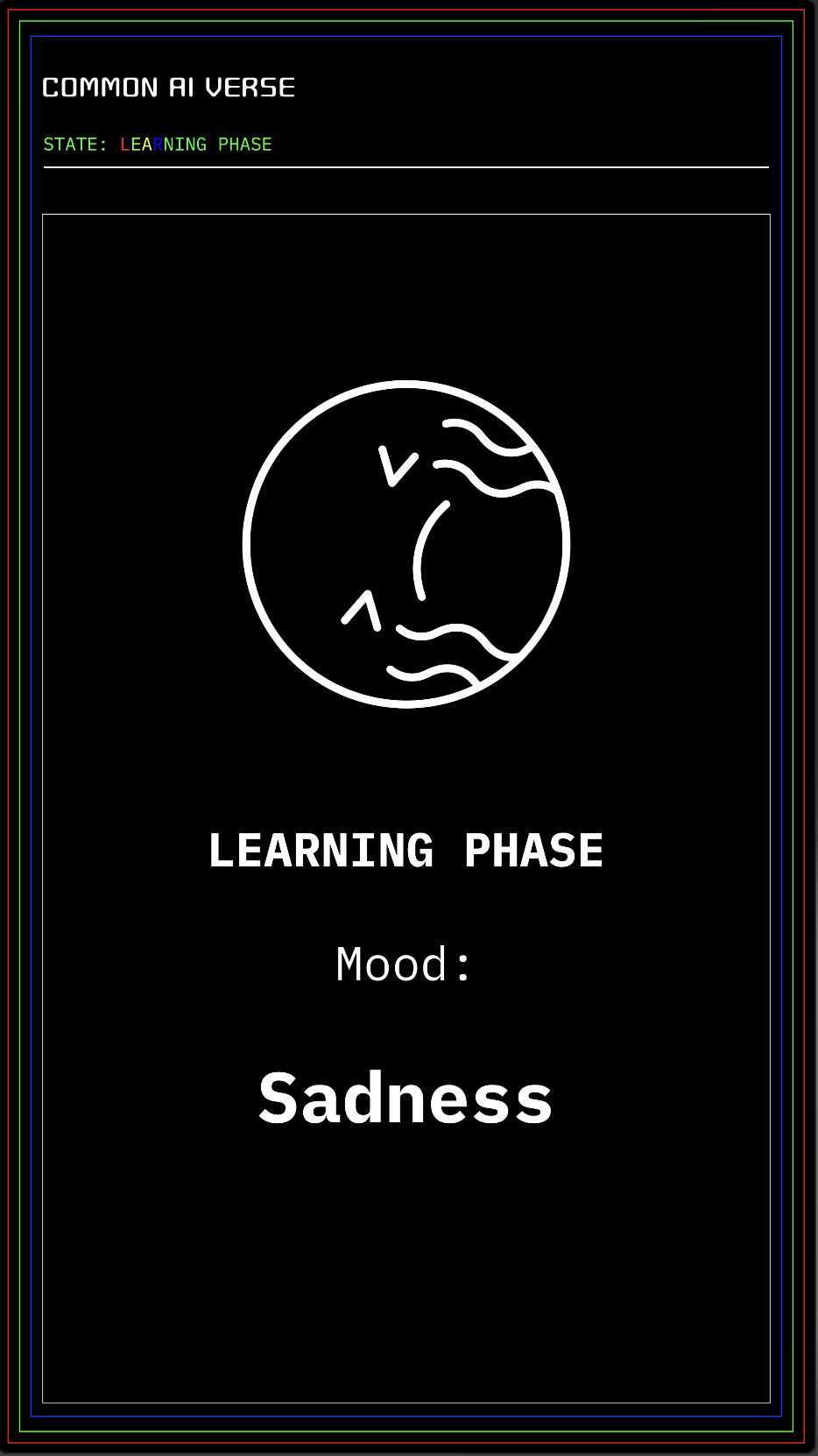}
    \caption{Initializing learning phase}
  \end{subfigure}

  \caption{Visual Feedback given the users while starting the process}.
\end{figure*}
\subsection{Audio}
The soundscape of the Commonaiverse installation was designed to dynamically interact with participants' movements and emotional expressions, inspired by established correlations between musical features and affective states \cite{gabrielsson2003emotional,hunter2010feelings}. This approach was rooted in existing literature on the psychophysiological and cognitive effects of musical elements—such as tempo, harmony, rhythm, and dynamics—on emotional perception.

To achieve a responsive and immersive auditory environment, the Max/MSP \cite{maxmsp} patch was integrated with TouchDesigner via Open Sound Control (OSC). Six speakers were strategically placed around the installation to create a spatial sound environment, enabling both an overarching auditory realm that unified the experience and individual soundscapes for each participant's interaction. The received data streams were mapped to modulate specific musical parameters dynamically:
\begin{itemize}
    \item \textbf{Tempo Adaptation:} Real-time movement velocity data influenced the musical tempo, with higher movement speeds triggering faster tempos and lower speeds generating slower tempos. This adaptation aligned the auditory environment with participants’ physical energy levels.
    \item \textbf{Harmonic Shifts:} The AI's analysis of participants' perceived emotional states guided transitions between major and minor tonalities, reflecting changes in the overall emotional atmosphere. For instance, a detected transition from joy to melancholy would result in a shift from a major to a minor key.
    \item \textbf{Rhythmic Complexity:} Movement patterns informed rhythmic variations, such as the introduction of syncopation or changes in beat regularity. Periods of high participant activity introduced more complex rhythmic patterns, while slower or steadier movements maintained a regular beat.

    \item \textbf{Dynamic Range:} The amplitude and intensity of participant movements were mapped to musical volume and dynamics. Bigger gestures resulted in louder, more intense soundscapes, while subdued movements created softer, more introspective auditory experiences.
\end{itemize}

\begin{figure}[ht]
  \centering
    \centering
    \includegraphics[width=1\linewidth]{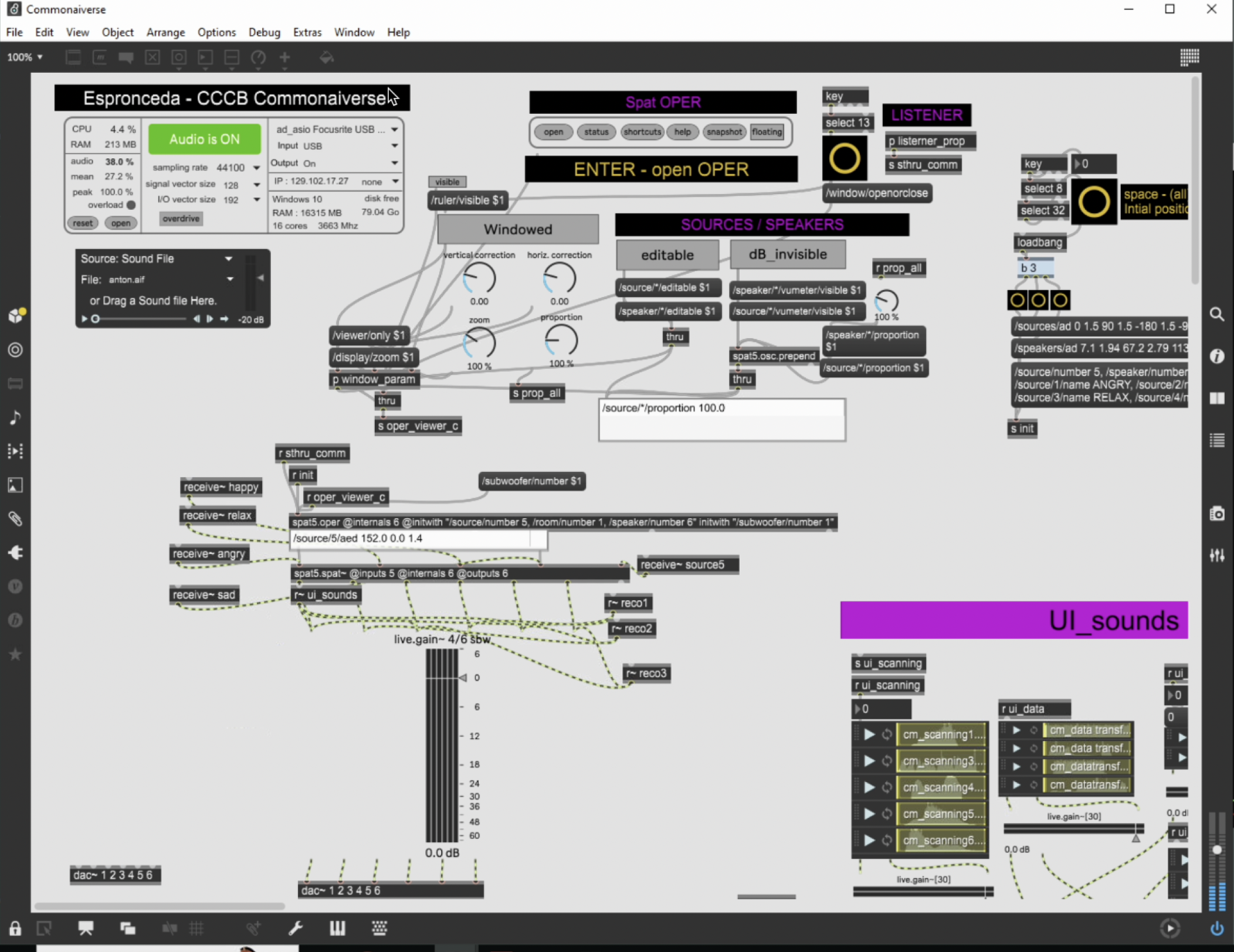}
    \caption{Audio Design of the Commonaiverse}
    \Description{Max/MSP screen showing different nodes for different emotion states and synths.}
\end{figure}
\subsection{Web3 Integration}
The information received throughout each session was conceptualized as a constellation in the cosmos of shared experiences. Thus, the Cosmos Phase of the installation served as a culmination of the interactions between participants and the AI, translating their collective movements and emotional states into a visually compelling and data-rich display. Upon completion, an on-screen visualization represented this data in an abstract, dynamic 3D form, symbolizing the emotional and physical interplay within the interactive space.

Each session’s data, including emotion levels, time factors, and movement patterns, was sent to a WebGL application. This real-time generator created unique crystal-like installations, where the size, creation time, rotation, and spatial relationships of each crystal piece were defined by the session’s data. These geometric structures and color-coded visuals were designed to reflect the emotional states detected by the AI, providing an artistic yet data-driven summary of the shared experience.

Participants could revisit this visualization through a QR code generated at the end of their session. The QR code encoded all the required information, enabling users to access and recreate their unique crystal installation via the WebGL application. This dual representation—combining the 3D visualization and session data metrics such as movement quantity, speed, and proximity—highlighted the collaborative and embodied nature of the interaction. More importantly, it offered participants a tangible way to reclaim ownership of their interaction data, creating a sense of continuity and personal connection beyond the installation.
\begin{figure}[ht]
  \centering
    \centering
    \includegraphics[width=0.8\linewidth]{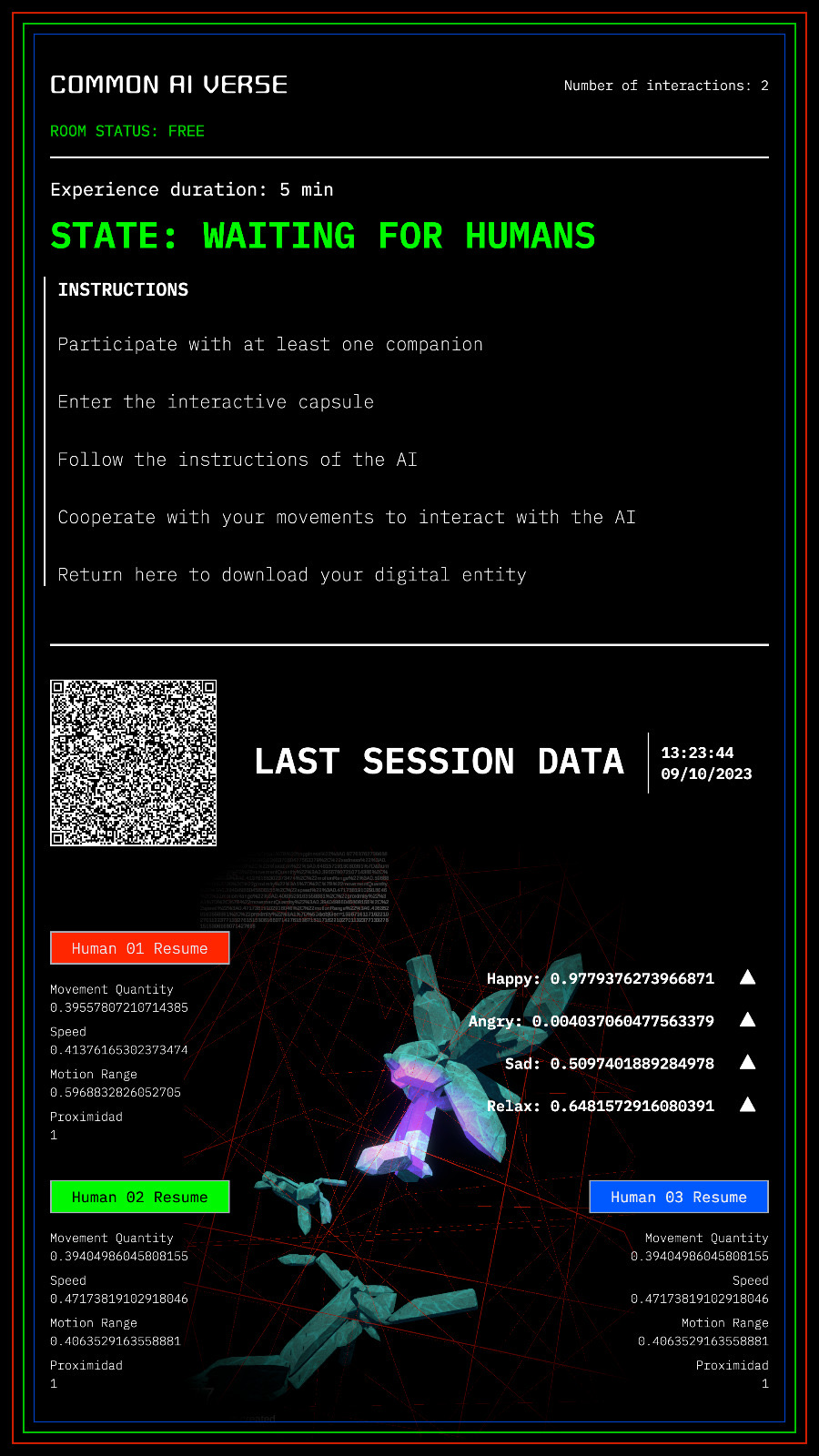}
    \caption{The Final State of the interaction, where people can access the data points and the constellation created from their interactions.}
    \Description{A screenshot showing the final state of the interaction, with a QR and information regarding user and collective data points.}
\end{figure}
\section{Critical Reflections}
Emotion recognition technologies are deeply embedded in the sociopolitical fabric of contemporary society. Their deployment in contexts such as workplace productivity tracking, consumer surveillance, and policing often reduces emotions to simplistic, monetizable metrics. This reductionist approach pervades many multimedia data pipelines, commodifying human affect and turning nuanced emotional states into data points for algorithmic optimization and profit generation. Commonaiverse resists this commodification by shifting focus from extraction to collaboration, inviting participants to co-create an embodied emotional lexicon with the AI.

This critique is particularly relevant given the biases inherent in existing emotion recognition systems. Many such systems are trained on datasets that encode cultural, racial, and gender biases, often privileging Western norms while marginalizing non-Western or minority expressions \cite{benjamin2019race,crawford2021atlas}. For instance, facial emotion recognition algorithms frequently misclassify expressions from individuals of non-Caucasian backgrounds \cite{raji2019actionable}. By requiring participants to teach the AI through embodied interactions, Commonaiverse decentralizes these normative assumptions, emphasizing diversity and rejecting universal emotional standards. This process aligns with critical calls to decolonize AI systems by prioritizing localized and situated knowledge \cite{irani2010postcolonial}.

While the multi-recommender system provides nuanced emotional detection through behavioral, contextual, and longitudinal analyses, its reliance on movement data introduces potential biases. For instance, privileging certain physical capabilities over others underscores the importance of inclusive design. Similarly, while the teaching phase enables participants to co-create an emotional lexicon, its adaptability across cultural contexts requires further exploration. In a broader multimedia sense, this highlights how model-building should account for diverse body types and movement repertoires to avoid reinforcing new forms of exclusion. Incorporating culturally specific movement patterns could enhance its universality and applicability.

The installation also critiques the pervasive surveillance logic that underpins many emotion recognition systems. When applied in policing or workplaces, these systems often function as tools for control, monitoring compliance or productivity without explicit consent \cite{zuboff2023age}. Commonaiverse counters this paradigm by embedding transparency and agency into its design. The Cosmos Phase, which allows participants to view and own their data, exemplifies a participatory ethos. This approach contrasts sharply with the extractive practices of commercial AI systems, presenting data as a resource for reflection and connection rather than exploitation.

Moreover, Commonaiverse interrogates the economic and political motivations driving the adoption of emotion recognition technologies. As emotions become commodities within surveillance capitalism, the human capacity for affective expression risks being dehumanized and instrumentalized \cite{couldry2020costs}. By reframing emotion recognition as an artistic and communal act, Commonaiverse reclaims affective expression as a domain for creativity and shared meaning-making. It invites participants to critically reflect on how their emotions are interpreted, valued, and potentially exploited within broader sociotechnical systems.

Finally, Commonaiverse contributes to speculative and critical HCI work that challenges conventional AI design paradigms. It aligns with research advocating for more ethical, inclusive, and situated technological approaches \cite{suchman2007human,dourish2011divining}. By positioning itself as both critique and alternative, the project fosters a dialogue about the politics of emotion in digital spaces. In doing so, it also prompts multimedia researchers to consider relationality, agency, and cultural specificity in designing the next generation of affective systems, rather than defaulting to reductive generalizations.

Despite its achievements, the project faces certain limitations. Real-time data processing occasionally misses subtle or overlapping emotional states, and performance can degrade under noisy or crowded conditions. Additionally, its reliance on visual and auditory feedback limits the depth of emotional interaction for some users. Addressing these limitations through iterative development will ensure that Commonaiverse remains inclusive and robust.

\section{Future Directions}
One of the immediate opportunities for Commonaiverse lies in its portability. Since the installation is not site-specific, adapting it as a traveling piece could provide a unique opportunity to gather data on diverse forms of bodily expressiveness across various cultures and environments. This traveling aspect would allow us to observe how different cultural contexts influence emotional communication and physical engagement, contributing to a richer understanding of embodied emotional expression and more inclusive AI modeling in multimedia contexts.

Future iterations could also benefit from moving beyond audiovisual stimuli to integrate additional sensory modalities, such as tactile feedback. Incorporating haptic elements would enhance the multisensory experience, allowing participants not only to see and hear but also to feel their interactions. This heightened level of sensory integration could deepen participants’ connection to the installation—and to each other—within immersive multimedia environments.

Moreover, Commonaiverse has the potential to inspire further applications in the realm of emotional interaction and embodiment. Using body tracking and emotional expression as a basis for therapeutic interventions, particularly in art therapy contexts, could foster emotional regulation and self-exploration. The expressive nature of full-body movement, combined with AI’s responsive feedback, provides fertile ground for emotional and psychological research within interactive multimedia frameworks.

Finally, integrating Web3 technologies for decentralized data storage could give participants ownership over their interaction data, fostering continuity and personal connection beyond the experience. This approach could also address broader concerns about data privacy and consent, aligning with the project’s ethos of collaboration and agency.

\section{Conclusion}
With Commonaiverse, our aim was to explore how AI can engage with human emotions in a way that feels embodied, dynamic, and collaborative. By integrating full-body movements, real-time motion tracking, and interactive multimedia feedback, we created a space where participants could express and interpret emotions together with the AI. Through its three-phase structure—Teaching, Exploration, and the Cosmos Phase—the installation bridges human expression and machine interpretation, emphasizing the social and relational aspects of emotional communication.

Throughout this process, we encountered and tackled challenges in modeling complex emotional states, balancing abstract artistic representations with accessibility, and generating collaboration in public, interactive settings. These challenges highlighted the importance of iterative design and reinforced our belief in the need for inclusivity and adaptability in designing emotion-focused technologies, moving away from static datasets and predefined categories.

Ultimately, Commonaiverse represents a vision of how AI can become an active participant in emotional exploration, not merely an observer. We hope this work inspires further experimentation at the intersection of technology, emotion, and embodiment, showing that AI can be a collaborator in understanding—and enhancing—human connection through the lens of multimedia innovation.

\begin{acks}

The work was exhibited in CCCB, Centre de Cultura Contemporània de Barcelona during the exhibition AI: Artificial Intelligence, between October 18th 2023 - March 17 2024, curated by Lluis Nacenta. 

This installation was a result of a collaboration among various artists: Solimán López, Esen K. Tütüncü, Kris Picher, Daniel Sabio and  Mohsen Hazrati. AI Integration was facilitated by IIIA-CSIC Barcelona team: Jordi Sabater, Joan Jené, Cristian Cozar and Lissette Lemus. We would like to thank Emmanuel Martinez for the web development, Carlos Reche for the audio design, and Marc Galvez for technical support. The project was produced by ESPRONCEDA: Institute of Art \& Culture with the support of the Department of Cultura of Catalunya and commissioned by Alejandro Martin with the executive production of Dr. Holger Sprengel. The early sketches, 3D models and the interaction visualizations were done by Solimán López. The image in Figure 5 was taken by Aleix Plademunt, Figure 6 by Marti E. Berenguer and Figure 7 by  Vitor Schietti.
\end{acks}


\bibliographystyle{ACM-Reference-Format}
\balance
\bibliography{sample-base}

\end{document}